# Effects of external magnetic field and hydrostatic pressure on magnetic and structural phase transitions in $Pr_{0.6}Sr_{0.4}MnO_3$

Amit Chanda and R. Mahendiran[1]

Department of Physics, National University of Singapore, 2 Science Drive 3,

Singapore -117551, Republic of Singapore.

**Abstract**

We investigate the effect of hydrostatic pressure on temperature dependence of magnetization and also the influence of magnetic field on linear thermal expansion in polycrystalline $Pr_{0.6}Sr_{0.4}MnO_3$, which is ferromagnetic at room temperature ($T_C$ = 305 K) but its magnetization undergoes an abrupt decrease at $T_S$ = 89 K within the ferromagnetic state. Normal and inverse magnetocaloric effects around $T_C$ and $T_S$, respectively, were reported earlier in this single phase compound [D. V. M. Repaka et al., J. Appl. Phys. 112, 123915 (2012)]. The thermal expansion shows an abrupt decrease at $T_S$ in zero magnetic field but it transforms into an abrupt increase at the same temperature under 7 T, which we interpret as the consequence of magnetic field-induced structural transition from a low-volume monoclinic ($I2/a$ symmetry) to a high volume orthorhombic ($Pnma$ symmetry) phase in corroboration with a published neutron diffraction study in zero magnetic field. While the external magnetic field does not change $T_S$, application of a hydrostatic pressure of $P$ = 1.16 GPa shifts the magnetic anomaly at $T_S$ towards high temperature. The pressure induced shift of the low-temperature anomaly ($\Delta T_S$ = 27 K) is nine-times more than that of the ferromagnetic Curie temperature ($\Delta T_C$ = 3K). Our results suggest that while hydrostatic pressure stabilizes the low temperature monoclinic phase at the expense of orthorhombic phase, magnetic field has an opposite effect.

---

[1] Author for correspondence (E-mail: Phyrm@nus.edu.sg)

## 1. Introduction

The perovskite manganite $Pr_{0.6}Sr_{0.4}MnO_3$ is a room temperature ferromagnet ($T_C$ = 305 K) exhibiting remarkable magnetocaloric, magnetothermoelectric, magnetoresistance, and magnetoimpedance effects around room temperature and thus has potential for applications.[1] An unusual property of this compound is that the temperature dependence of field-cooled magnetization $M(T)$ under low magnetic fields decreases abruptly at $T_S$ = 89 K upon cooling, much below the ferromagnetic transition and the anomaly vanishes as the magnetic field increases to more than 1 Tesla.[1] A consequence of this low-temperature magnetic anomaly is the inverse magnetocaloric effect (MCE) in which magnetic entropy change is positive ($\Delta S_m = S_m(H)-S_m(H=0)$) in presence of external magnetic fields, which is contrary to negative $\Delta S_m$ (normal MCE) usually encountered in ferromagnets and paramagnets. The coexistence of both normal and inverse MCE in a single phase material is not very common and it is interesting because magnetic cooling can be achieved by adiabatic demagnetization as well as adiabatic magnetization processes in different temperature regimes, which will enable extension of the working temperature range for magnetic refrigeration. Inverse MCE has been reported in some ferromagnetic Heusler alloys undergoing shear driven austentite-martensite transition, where the magnitude of inverse MCE dominates over that of the normal MCE.[2] Interestingly, the abrupt decrease of magnetization within the ferromagnetic state of $Pr_{0.6}Sr_{0.4}MnO_3$ sample is not due to onset of antiferromagnetic transition but caused by a structural change from orthorhombic (*Pnma* space group) to monoclinic (*I2/a* space group) phase as revealed by a neutron diffraction work on the same composition reported several years ago.[3] Heat capacity also showed an anomaly at this temperature[4], however, electrical resistivity did not register any anomaly.[1]

The electrical behavior of $Pr_{0.6}Sr_{0.4}MnO_3$ is in contrast to $Pr_{0.5}Sr_{0.5}MnO_3$ which also shows a ferromagnetic transition ($T_C = 265$ K) and abrupt decrease of $M(T)$ at low temperature ($\approx$ 135 K) which triggers a jump in resistivity at the same temperature. The low-temperature magnetic anomaly in $Pr_{0.5}Sr_{0.5}MnO_3$ is the consequence of a ferromagnetic (FM) to a layered antiferromagnetic (A-type AFM) transition that is shown to be simultaneously accompanied by a tetragonal (*I4/mcm*) to orthorhombic (*Fmmm*) first-order structural transition.[5] It still remains a puzzle whether the FM-AFM transition is triggered by the structural phase transition or the structural transition changes the nature of spin ordering. Application of hydrostatic pressure is a clean experimental technique to probe magnetic phase transitions without introducing chemical disorder and extra charge carriers. Hydrostatic pressure affects structural parameters, Mn-O-Mn bond angle and Mn-O distance which in turn modify exchange interactions between Mn ions. In ferromagnetic manganites, application of a moderate hydrostatic pressure ($P < 2$ GPa) usually increases the ferromagnetic Curie temperature ($T_C$) and it is attributed to widening of conduction bandwidth and reduction of electron-phonon coupling with pressure.[6] A recent review by Markovich *et al*. summarizes the pressure effect on magnetic phase transitions in a wide range of manganites.[7] Strangely, magnetic properties of $Pr_{0.5}Sr_{0.5}MnO_3$ under hydrostatic pressure has not been reported until now. However, the effect of hydrostatic pressure on electrical resistivity in polycrystalline $Pr_{0.5}Sr_{0.5}MnO_3$ sample was studied by Rueckert *et al*.,[8] who found that resistivity in the AFM phase decreased with increasing pressure but the onset of AFM transition temperature (Neel temperature, $T_N$, identified from the onset temperature for the abrupt jump in resistivity) shifted to high temperature. On the other hand, $T_C$ decreased with increasing hydrostatic pressure. The application of an external magnetic field, in contrast to hydrostatic pressure, enhanced $T_C$ and lowered $T_N$ in the same composition. Earlier study on the pressure dependence of magnetization under moderate pressure ($P < 2$ GPa) in $Pr_{1-x}Sr_xMnO_3$ series was confined to ferromagnetic samples of compositions $x = 0.22 - 0.26$, which did not show structural transitions below $T_C$ [9] and the external pressure was found to enhance $T_C$ in contrast to the suppression of $T_C$ reported in $x = 0.5$.

The absence of AFM transition at low temperatures in $Pr_{0.6}Sr_{0.4}MnO_3$ is an advantage compared to $Pr_{0.5}Sr_{0.5}MnO_3$. In the above context, investigation of the hydrostatic pressure on the low-temperature magnetic anomaly triggered only by structural transitions in $Pr_{0.6}Sr_{0.4}MnO_3$ is interesting. In addition to investigating the pressure dependence of magnetization, we also report thermal expansion without and with a magnetic field in this compound for the first time. Our results show that effects of external magnetic field and hydrostatic pressure on structural phase transition are different in this material.

## 2. Experimental details

Polycrystalline sample of $Pr_{0.6}Sr_{0.4}MnO_3$ was prepared from stoichiometric mixtures of high purity $Pr_6O_{11}$, $SrCO_3$, and $Mn_2O_3$ by the conventional solid-state reaction route. After two intermediate grindings and heating at 1000°C for 12 h and 1100°C for 24 h in air, the mixture was pressed into pellets and sintered at 1200°C for 48 h. The magnetization (M) of the sample under ambient and hydrostatic pressure was measured using a vibrating sample magnetometer (VSM) incorporated to a commercial superconducting cryostat (Physical Property Measurement System (PPMS). A piston type pressure cell assembly made up of Cu-Be alloy was used for the pressure dependent magnetization measurement. The sample was first fully emerged in Dauphine oil (acts as a pressure transmitting media) inside a Teflon tube and then encapsulated using Teflon caps from both ends of the tube. The encapsulated Teflon tube was inserted within a sample holder on the both side of which two pistons were attached. By tightening the screws attached to the pistons, a longitudinal force can be applied to the pistons and hence, on the Teflon caps from both the sides. This force transmits through the Dauphine oil and exerts pressure on the sample. The entire pressure cell assembly was attached to the VSM sample rod and inserted inside the PPMS. Standard Pb sample was used for the calibration purpose by monitoring the pressure dependence of its superconducting transition temperature. Linear thermal expansion of the sample was measured using a miniature dilatometer probe attached to the PPMS. Four probe electrical resistance of the sample was also measured in PPMS.

## 3. Results

Fig. 1 shows the room temperature X-ray diffraction pattern and the corresponding FullProf profile fitting of the sample $Pr_{0.6}Sr_{0.4}MnO_3$ (PSMO), which reveals that the sample is in single phase and has orthorhombic crystal structure with space groups *Pnma* at room temperature. The values of the lattice parameters *a*, *b* and *c* are 5.484, 7.648 and 5.448 Å, respectively.

Figs. 2(a) shows the temperature dependence of field-cooled (FC) magnetization (*M*) of the PSMO sample under three different magnetic fields ($\mu_0H$ = 0.01, 0.1 and 7T) in the absence of external pressure (*P* = 0 GPa) in the temperature range *T* = 350-10 K. Fig. 2(b) shows the temperature derivative of magnetization, *dM/dT*, curves for all the three magnetic fields. The rapid increase of M(T) signals the onset of ferromagnetic transition. The ferromagnetic Curie temperature ($T_C$) identified from the minimum of $dM/dT$ for $\mu_0H$ = 0.01 T is 305 K. While $T_C$ remains unaltered for 0.1 T, it shifts to a higher temperature under 7 T. Within the ferromagnetic state, M(T) shows downward step at $T_S$ = 89 K for $\mu_0H$ = 0.01 and 0.1 T and hence, $dM/dT$ exhibits a peak at $T_S$. The step disappears in M(T) under 7 T. The *dc* resistivity $\rho(T)$ in zero magnetic field shows a peak at $T_C$ and no anomaly is detected in $\rho(T)$ around $T_S$ (see fig. 2(c)). In the presence of 7 T magnetic field, resistivity peak shifts to 325 K and the magnitude of resistivity decreases in the entire temperature range measured.

Fig. 3(a) displays the temperature dependence of linear thermal expansion expressed as relative length change $\Delta L//L$ = [L(T)-L(T = 315 K)]/L(T = 315 K) in the absence of external magnetic field ($\mu_0H$ = 0 T) and in the presence of $\mu_0H$ = 7 T. The strain is measured in the direction of the applied magnetic field. We have normalized the length change with respect to its length at 315 K where the sample is in paramagnetic state. Our capacitance dilatometer cannot be operated above 320 K. $\Delta L/L$ decreases with decreasing temperature implying contraction of the sample length upon cooling. While $\Delta L/L$ in zero field barely shows anomaly at $T_C$, it exhibits a steplike decrease around $T_S$ = 89 K. The application of a magnetic field (7T) dramatically reduces the thermal contraction at and below $T_S$ compared to its influence above $T_S$. At 10 K, $\Delta L/L$ increases by 5 x$10^{-3}$ under 7 T magnetic field and the downward step at $T_S$ is replaced by an upward step. Thus, a positive magnetostriction occurs. For clarity, $\Delta L/L$ as function of temperature for different magnetic fields between 0 and 7 T are shown in the inset of Fig. 3(a) which indicates gradual weakening of the downward step at $T_S$ with increasing strength of the magnetic field and its transformation to a small upward step at $T_S$ under 5 T magnetic field. As shown in the inset of Fig. 3(b), $\delta(\Delta L/L)_{7T}$ which is the difference in $\Delta L/L$ between 0 and 7T magnetic fields shows a sharp rise at $T_S$. We plot the coefficient of thermal expansion, $\alpha(T) = (1/L)d(\Delta L)/dT$ for $H = 0$ T and 7 T in the main panel of Fig. 3(b). Although $\alpha(T)$ in zero field shows a weak λ-like anomaly around $T_C$, it is masked by a dominant positive peak that appears at $T_S$. In the presence of 7 T magnetic field, a prominent dip occurs in $\alpha(T)$ at $T_S$. Since the peak and the dip occurs at the same temperature, $T_S$ is not shifted by the application of magnetic field.

Fig. 4(a) shows the impact of the hydrostatic pressure ($P = 1.16$ GPa) on $M(T)$ of PSMO for $H = 1$ kOe along with the ambient pressure ($P = 0$ GPa) data. Fig. 4(b) shows the temperature derivative of magnetization, $dM/dT$. It is seen that hydrostatic pressure increases the ferromagnetic transition temperature ($\Delta T_C = +3$). However, the impact of hydrostatic pressure on the low temperature anomaly is more remarkable compared to the PM-FM transition. The applied pressure causes a large upward shift of $T_S$ ($\Delta T_S = 27$ K) which can be clearly seen in $dM/dT$ curves. The values of magnetization under pressure are lower than the zero pressure values below $T_C$ and $M(T)$ under 1.16 GPa pressure shows a tendency to decrease rapidly with temperature below $T_C$ compared to the zero pressure $M(T)$.

The main panel of Fig. 5 shows $M(H)$ hysteresis loop of PSMO at $T = 5$ K recorded for both $P = 0$ and 1.16 GPa. The high field magnetization decreases slightly under pressure which is not in disagreement with low-field $M(T)$ behavior though the decrease is not as large as the $M(T)$ curve recorded under 0.1 T depicts. At the highest field of 7 T, $M_S$ of the PSMO samples drops from 3.58 $\mu_B$/f.u. at $P = 0$ to 3.47 $\mu_B$/f.u. at $P = 1.16$ GPa. The insets of the Fig. 5 depicts enlarged view of the hysteresis loop for PSMO. The coercivity ($H_C$) shrinks from 383 Oe to 295 Oe under the application of 1.16 GPa pressure.

## 4. Discussion

The results presented above indicate that the application of hydrostatic pressure affects both the ferromagnetic Curie temperature and the low temperature magnetic anomaly. The shift of $T_S$ is higher than $T_C$ itself under hydrostatic pressure. In contrast to the influence of hydrostatic pressure, the low-temperature magnetic anomaly is not shifted by the application of magnetic field. Thermal expansion shows a pronounced anomaly at $T_S$ even at the highest magnetic field of 7 T although the anomaly is absent in $M(T)$ measured at 7 T.

Let us first give qualitative explanations for the observed thermal expansion behavior and the effect of magnetic field on the thermal expansion. The abrupt decrease of $\Delta L/L$ around 89 K in PSMO where the low-field magnetization also decreases abruptly underscores a close link between magnetism and structural changes. A structural transition from orthorhombic (*Pnma*) at high temperature to monoclinic (*I2/a*) at low temperature around $T_S$ in same nominal composition was confirmed by the neutron diffraction work of Ritter *et al.*[3] First-order nature of this structural transition was also confirmed by a hysteresis in field-cooled and field-warm magnetizations and differential thermal analysis data of Maheswar *et al.*[1] According to Ritter *et al.*,[3] the structure of $Pr_{0.6}Sr_{0.4}MnO_3$ does not transform completely from orthorhombic to monoclinic at $T_S$ (= 105 K in their sample) but only 38% volume fraction (VF) of the *Pnma* phase transforms into monoclinic phase at 90 K, and the VF transforming into *I2/a* phase increases with lowering temperature. Maximum transformation occurs between 90 and 60 K, from 38% at 90 K to 87% at 60 K, and below 60 K, the rate of transformation slows down.

While Ritter *al.*, reported the presence of 88% VF of *I2/a* phase at 1.6 K, Boujelben *et al.*,[10] reported 73% VF of *I2/a* phase at 10 K. Hence, it seems that the ratio of these two phases below $T_S$ is dependent on sample synthesis conditions which can introduce a slight non-stoichiometry. The observed behavior of thermal expansion in non zero magnetic fields may be an indication of a volume change rather than change in length alone. To measure magnetovolume change, magnetostriction parallel and perpendicular the magnetic field directions need to be measured, however, it is not possible to measure magnetostriction perpendicular to the magnetic field direction in our capacitance dilatometer set up. We suppose that the abrupt contraction of length at $T_S$ implies a sudden decrease in the unit cell volume at $T_S$ because the low-temperature structural transition is first order as indicated by previous works. The low-temperature monoclinic phase has lower volume than the high-temperature orthorhombic phase. Since both these phases are ferromagnetic, the decrease of magnetization at $T_S$ is possibly caused by a change in the direction of easy axis during the structural transition. According to Ritter *et al.*,[3] the magnetic structure is a simple ferromagnetic type with the moment aligned in direction of longest axis (*a* axis in *I2/a* and *b* axis in *Pnma*) in zero magnetic field. When 7 T magnetic field is applied in the paramagnetic phase and the sample is cooled to 10 K, the easy axis of magnetization in the low temperature phase tends to rotate towards the direction of the applied field and in this process the low-volume monoclinic phase undergoes a structural transformation into a high-volume phase. The high volume phase may be orthorhombic. The volume fraction of the transformed phase increases with increasing strength of the cooling magnetic field. This explains the enormous change in $\Delta L/L = 5 \times 10^{-3}$ found at 10 K for 7 T. The upward jump in $\Delta L/L$ at $T_S$ for $H = 7$ T suggests that the transformed phase at 7T has larger volume than that of the orthorhombic phase existing just above $T_S$. The absence of anomaly in $M(T)$ under 7 T implies that the Mn moments in the new phase are mostly aligned in the field direction. Field induced structural transition was also proved by neutron diffraction under magnetic field in the charge-ordered antiferromagnetic $Nd_{0.5}Sr_{0.5}MnO_3$, in which the low-temperature monoclinic phase was transformed completely into orthorhombic phase under 6 T magnetic field along with destruction of charge ordered (CE type) antiferromagnetic state and emergence of ferromagnetic charge delocalized state.[11] However, in ferromagnetic manganites like Y-substituted $La_{0.67}Ca_{0.33}MnO_3$, the volume magnetostriction is large just above $T_C$ and is negligibly small at 10 K.[12] The origin of volume magnetostriction (volume decreases under external magnetic field) above $T_C$ is related to the formation of magnetic polarons (lattice polarons with a magnetic cloud) in zero field and suppression of the magnetic polarons with increasing strength of magnetic field. On the other hand, the volume magnetostriction is largest at the lowest temperature in our sample.

Now let us consider the effect of hydrostatic pressure on magnetization. The $T_C$ of PSMO sample increases by 3 K under the hydrostatic pressure of $P = 1.16$ GPa. In general, $T_C$ of the ferromagnetic manganites shifts with increasing pressure for moderate pressures ($P < 2$ GPa) and it is a well research topic now. In manganites, the strength of the ferromagnetic double exchange interaction depends on the $e_g$-electron bandwidth $W = \cos^2\phi / (d_{Mn-O})^{3.5}$, where, $\phi$ is the Mn-O-Mn bond angle and $d_{Mn-O}$ represents the Mn-O bond length. The ferromagnetic Curie temperature is proportional to the $e_g$-electron conduction band width ($T_C = \beta W$) where $\beta$ is a constant). Hence, an increase in $\phi$ and/or contraction of Mn-O bond under hydrostatic pressure leads to enhancement in $W$ and thereby, increase in $T_C$. In our sample, the low-temperature magnetic anomaly and thus the structural transition temperature shifts up rapidly with applied pressure compared to the ferromagnetic Curie temperature itself. Hence, the hydrostatic pressure tends to stabilize and expands the monoclinic phase at the expense of orthorhombic phase. Why is it so? A likely possibility is that that hydrostatic pressure uniaxially compresses the $MnO_6$ octahedra, which will lower the energy of $e_g$: $d_{x2-y2}$ orbital state compared to $e_g$: $d_{3z2-r2}$ orbitals. It is worth to point out a few neutron diffraction studies done under high pressure in compounds with higher Sr content than our sample. D. P. Kozlenko et al.[13] investigated high pressure effect on crystal and magnetic structure of $Pr_{0.52}Sr_{0.48}MnO_3$. This sample is ferromagnetic below $T_C$ = 270 K and possesses tetragonal (space group *I4/mcm*) structure in ambient pressure. For $P \geq 2$ GPa, the ferromagnetic phase transforms into A-type antiferromagnetic phase ($T_N \approx 250$ K) and it is shown to be accompanied by tetragonal (*I4/mcm*) to orthorhombic (*Fmmm*) structural transition. No magnetization, transport or thermal expansion data were provided in that report. In the magnetic phase diagram of $Pr_{1-x}Sr_xMnO_3$, A-type antiferromagnetic phase is present even in ambient pressure at low temperatures in a narrow composition range (0.5 ≤ *x* ≤ 0.56). $Pr_{0.46}Sr_{0.56}MnO_3$ undergoes a simultaneous paramagnetic to A-type antiferromagnetic transition and tetragonal (*I4/mcm*)-orthorhombic (*Fmmm*) transition at $T_N$ = 215 K in ambient pressure.[14] However, application of pressure more than 2 GPa to this sample induces a C-type antiferromagnetic phase (tetragonal structure) with $T_N$ = 120 K. Both A-type orthorhombic and C-type tetragonal phases coexist down to the lowest temperature (*T* = 16 K) measured under hydrostatic pressure. These results indicate that a hydrostatic pressure more than 2 GPa create new structural and antiferromagnetic phases and these phases can coexist with the ambient pressure high temperature magnetic phases. However, the applied pressure is not very high (*P* = 1.16 GPa) in our sample and it is unlikely that a long range A-type antiferromagnetic order is established in our sample. *M(H)* isotherm at 10 K shows ferromagnetic behavior under 1.16 GPa though with a reduced high field magnetization and coercive field rather than antiferromagnetic behavior. However, the $M(T)$ curve under 1.16 GPa in our sample shows a rapid decrease with decreasing temperature below $T_C$ compared to the ambient pressure $M(T)$ curve, which may be considered as an indication of the development of short-range antiferromagnetic correlations among Mn spins in predominantly ferromagnetic matrix. Hence, it appears that structural transition and accompanying $e_g$-orbital polarization are driving A-type antiferromagnetism in half-doped $Pr_{0.5}Sr_{0.5}MnO_3$. Crystallographic and magnetic structural information by neutron diffraction under high pressure will be useful to understand our results.

**Summary**

In summary, we have investigated magnetization in ambient pressure as well as under hydrostatic pressure, $P =1.16$ GPa and thermal expansion in zero and non-zero magnetic fields in the room temperature ferromagnet $Pr_{0.6}Sr_{0.4}MnO_3$. Thermal expansion showed an abrupt decrease at $T_S = 89$ K ($<< T_C = 305$ K) where the field cooled magnetization also decreased abruptly. The application of magnetic field led to a huge positive magnetostriction effect below $T_S$, which we attributed to the field-induced monoclinic to orthorhombic structural transition. In contrast to the effect of external magnetic field, hydrostatic pressure stabilized the low-temperature monoclinic (low-volume) phase at the expense of the high-temperature orthorhombic (high-volume) phase. It will be interesting to investigate whether the inverse magnetocaloric effect around $T_S$ is enhanced by the application of hydrostatic pressure.

**Acknowledgements**: R. M. thanks the Ministry of Education, Singapore for supporting this work through a Tier-2 grant (Grant no. R144-000-373-112/MOE-T2-2-147).

**Figure captions:**

**FIG. 1.** X-ray diffraction patterns and corresponding FullProf profile fittings of $Pr_{0.6}Sr_{0.4}MnO_3$.

**FIG. 2**(a) The temperature dependent field-cooled (FC) magnetization $M(T)$ under 100 Oe, 1 kOe and 7 T magnetic fields for $Pr_{0.6}Sr_{0.4}MnO_3$, (b) corresponding $dM/dT$ curves and $Pr_{0.6}Sr_{0.4}CoO_3$, respectively and (c) temperature dependent resistivity $\rho(T)$ under 0 and 7 T magnetic fields.

**FIG. 3**(a) The temperature dependent linear thermal expansion $\Delta L/L$ under 0 and 7 T magnetic fields in $Pr_{0.6}Sr_{0.4}MnO_3$ sample. Inset of (a) shows $\Delta L/L$ under different magnetic fields. (b) Thermal expansion coefficient $\alpha = (1/L)d(\Delta L)/dT$ and the inset shows the difference in $\Delta L/L$ under 7 and 0 T magnetic fields as a function of temperature.

**FIG. 4**(a) $M(T)$ at 1 kOe magnetic field and under $P = 0$ and 1.16 GPa hydrostatic pressure for $Pr_{0.6}Sr_{0.4}MnO_3$ and corresponding $dM/dT$ for $Pr_{0.6}Sr_{0.4}MnO_3$. Inset of (a) shows schematic structure of $MnO_6$ octahedra under ambient pressure and inset of (b) shows schematic structure of $MnO_6$ octahedra under $P = 1.16$ GPa.

**FIG. 5** The magnetic field dependent magnetization, $M(H)$ at 5 K under $P = 0$ and 1.16 GPa hydrostatic pressure for $Pr_{0.6}Sr_{0.4}MnO_3$, inset shows enlarged view of the $M(H)$.

**References:**

[1] D. V. Maheswar Repaka, T. Tripathi, M. Aparnadevi, and R. Mahendiran, J. Appl. Phys. **112**, 123915 (2012).

[2] M. Khan, N. Ali, and S. Stadler, J. Appl. Phys. **101**, 053919 (2007); T. Krenke, E. Duman, M. Acet, E. F. Wassermann, X. Moya, L. Manosa and A. Planes, Nature Mater. **4**, 450 (2005).

[3] C. Ritter, P.G. Radaelli, M.R. Lees, J. Baratt, G. Balakrishnan, D. McK Paul, J. Solid State Chem., **121**, 276 (1996).

[4] S. Rosler, H. S. Nair, U. K. Rosler, C.M.N. Kumar, S. Elizabeth and S. Wirth, Phys. Rev. B **84**, 184422 (2011).

[5] F. Damay, C. Martin, M. Hervieu, A. Maignan, B. Raveau, G. Andre, and F. Bouree, J. Magn. Magn. Mater. **184**, 71 (1998).

[6] V. Laukhin, J. Fontcuberta, J. L. García-Muñoz and X. Obradors, Phys. Rev. B **56**, R10009 (1997).

[7] V. Markovich, A. Wisniewski and H. Szymczak, Chapter 1, in Handbook of Magnetic Materials, vol **22** (2014) (Ed. K. H. J. Buschow), Elsevier, Amsterdam.

[8] F. J. Rueckert, M. Steiger, B. K. Davis, T. Huynh, J. J. Neumeier, and M. S. Torikachvili, Phys. Rev. B **77**, 064403 (2008).

[9] V. Markovich, I. Fita, R. Puzniak, A. Wisniewski, K. Suzuki, J. W. Cochrane, Y. Yuzhelevskii, Ya. M. Mukovskii, and G. Gorodetsky, Phys. Rev.B **71**, 224409 (2005).

[10] W. Boujelben, M. Ellouze, A. Cheikh-Rouhou, J. Pierre, Q. Cai, W. Yelon, K. Shimizu, and C. Dubourdieu, J. Alloys and Comp. **334**, 1 (2002).

[11] C. Ritter, R. Mahendiran, M.R. Ibarra, L. Morellon, A. Maignan, B. Raveau, and C. N. R. Rao, Physical Review B **61**, R9229 (2000).

[12] J. M. de Teresa, M. R. Ibarra, P. A. Algarabel, C. Ritter, C. Marquina, J. Blasco, J. Garcia, A. del Moral, and Z. Arnold, M. R. Ibarra, Nature (London) **386**, 256 (1997).

[13] D. P. Kozlenko, L. S. Dubrovinski, Z. Jirak, B. N. Savenko, C. Martin, and S. Vratislav, Phys. Rev. B **76**, 094408 (2007).

[14] D. P. Kozlenko, V. P. Glazkov, Z. Jirak, and B. N. Savenko, J. Phys.:Condens. Matter. **16**, 2381 (2014).

**Figures**

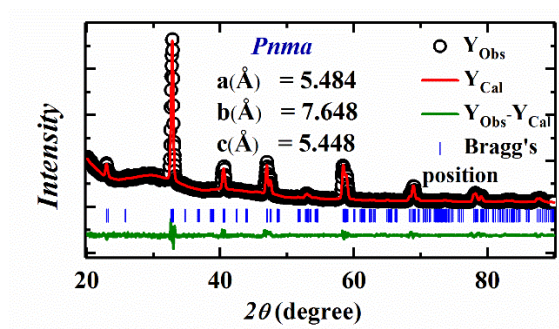

**FIG. 1.** X-ray diffraction patterns and corresponding FullProf profile fittings of $Pr_{0.6}Sr_{0.4}MnO_3$.

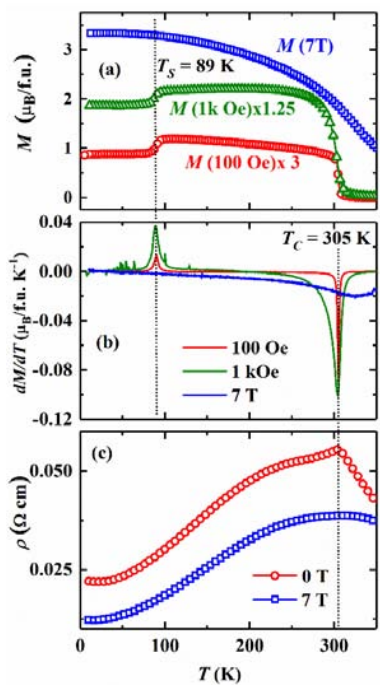

**FIG. 2**(a) The temperature dependent field-cooled (FC) magnetization $M(T)$ under 100 Oe, 1 kOe and 7 T magnetic fields for $Pr_{0.6}Sr_{0.4}MnO_3$, (b) corresponding $dM/dT$ curves and $Pr_{0.6}Sr_{0.4}CoO_3$, respectively and (c) temperature dependent resistivity $\rho(T)$ under 0 and 7 T magnetic fields.

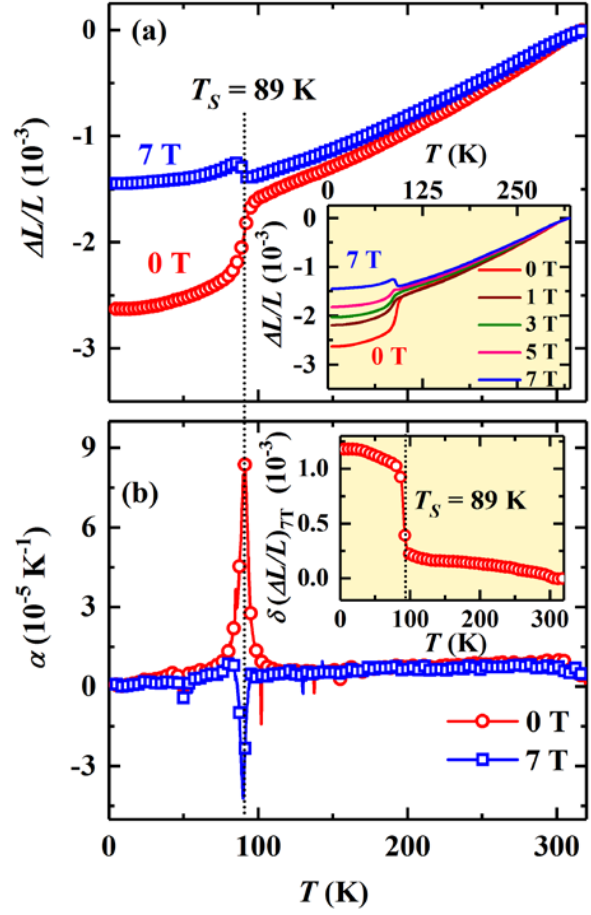

**FIG. 3**(a) The temperature dependent linear thermal expansion $\Delta L/L$ under 0 and 7 T magnetic fields in $Pr_{0.6}Sr_{0.4}MnO_3$ sample. Inset of (a) shows $\Delta L/L$ under different magnetic fields. (b) Thermal expansion coefficient $\alpha = (1/L)d(\Delta L)/dT$ and the inset shows the difference in $\Delta L/L$ under 7 and 0 T magnetic fields as a function of temperature.

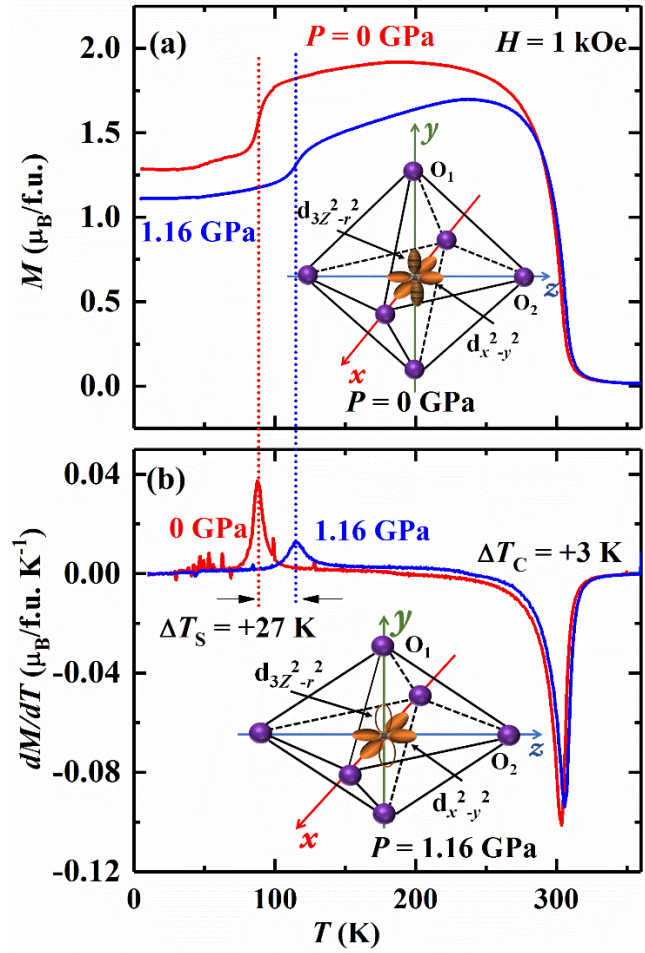

**FIG. 4**(a) $M(T)$ at 1 kOe magnetic field and under $P = 0$ and 1.16 GPa hydrostatic pressure for $Pr_{0.6}Sr_{0.4}MnO_3$ and corresponding $dM/dT$ for $Pr_{0.6}Sr_{0.4}MnO_3$. Inset of (a) shows schematic structure of $MnO_6$ octahedra under ambient pressure and inset of (b) shows schematic structure of $MnO_6$ octahedra under $P = 1.16$ GPa.

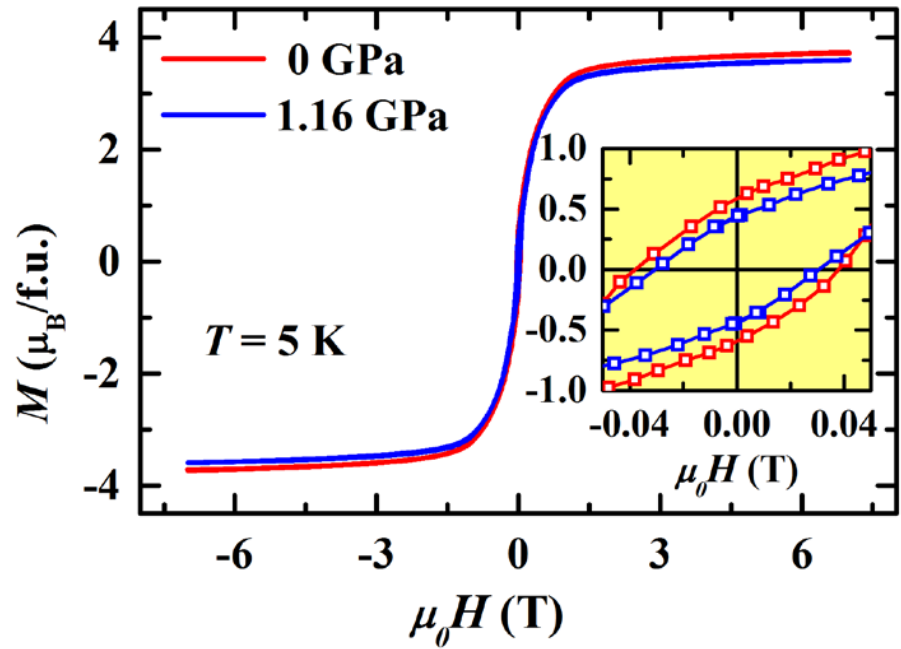

**FIG. 5** The magnetic field dependent magnetization, $M(H)$ at 5 K under $P = 0$ and 1.16 GPa hydrostatic pressure for $Pr_{0.6}Sr_{0.4}MnO_3$, inset shows enlarged view of the $M(H)$.